\documentclass[12pt,reqno]{amsart}
\usepackage{latexsym}

\newcommand{\K}{{\mathbf K}}
\newcommand{\tr}{{\operatorname{Tr}}}
\newcommand{\sym}{{\operatorname{Sym}}}
\newcommand{\HS}{{\operatorname{HS}}}
\newcommand{\crit}{{\operatorname {crit}}}
\newcommand{\inv}{^{-1}}
\newcommand{\kahler}{K\"ahler }
\newcommand{\wh}{\widehat}
\newcommand{\PP}{{\mathbb P}}
\newcommand{\R}{{\mathbb R}}
\newcommand{\C}{{\mathbb C}}

\newcommand{\CP}{\C\PP}
\newcommand{\U}{{\rm U}}
\newcommand{\half}{{\frac{1}{2}}}
\newcommand{\vol}{{\operatorname{Vol}}}
\newcommand{\hcal}{\mathcal{H}}
\newcommand{\ical}{\mathcal{I}}

\newcommand{\ncal}{\mathcal{N}}
\newcommand{\ocal}{\mathcal{O}}
\newcommand{\al}{\alpha}
\newcommand{\be}{\beta}
\newcommand{\ga}{\gamma}
\newcommand{\Ga}{{\rm \Gamma}}
\newcommand{\la}{\lambda}
\newcommand{\ep}{\varepsilon}
\newcommand{\de}{\delta}
\newcommand{\De}{{\rm \Delta}}
\newcommand{\Th}{{\rm \Theta}}
\newcommand{\Ph}{{\rm \Phi}}
\newcommand{\X}{{\rm \Xi}}
\newcommand{\om}{\omega}

\newtheorem{theo}{{\sc Theorem}}[section]
\newtheorem{cor}[theo]{{\sc Corollary}}
\newtheorem{lem}[theo]{{\sc Lemma}}

\title[asymptotics and dimensional dependence of critical points] {Asymptotics and dimensional dependence of the number of critical points of random holomorphic sections}
\author{Benjamin Baugher}
\address{Department of Mathematics, Johns Hopkins University, Baltimore, MD 21218, USA} 
\email{bbaugher@math.jhu.edu}
\date{Mar. 26, 2007}

\begin{document}

\begin{abstract}

We prove two conjectures from \cite{DSZ2,DSZ3} concerning the expected number of critical points of random holomorphic sections of a positive line bundle.  
We show that, on average, the critical points of minimal Morse index are the most plentiful for holomorphic sections of $\ocal (N) \to \CP^m$ and, in an asymptotic sense, for those of line bundles over general \kahler manifolds.  We calculate the expected number of these critical points for the respective cases and use these to obtain growth rates and asymptotic bounds for the total expected number of critical points in these cases.  This line of research was motivated by landscape problems in string theory and spin glasses.

\end{abstract}

\maketitle

\section{Introduction}\label{A1}

In the series of articles \cite{DSZ1,DSZ2,DSZ3}, the authors have been studying the statistics of critical points of Gaussian random holomorphic sections and their application to the vacuum selection problem in string theory.  The purpose of this article is to prove two conjectures from these papers. 

In \cite{DSZ2,DSZ3} it was informally conjectured that the expected number $\ncal^\crit_{N,h}(\CP^m)$ of critical points of random holomorphic sections of $\ocal (N) \to \CP^m$ grows exponentially with the dimension.  This was based on a conjectured formula for the expected number of critical points of minimal Morse index and the evidence from calculations in small dimensions that the expected number $\ncal^\crit_{N,q,h}(\CP^m)$ of critical points of Morse index $q$ decreased as $q$ increased.  In \cite[Sec. 7.3]{DSZ3} this conjectured growth rate was used as a basis for the heuristic estimate of the growth rate for the expected density of vacua in string/M theory.  It was also noted that it is consistent with the analogous estimates of the growth rate of the number of metastable states of spin glasses \cite {F}.  In this paper we show that this conjecture is indeed true by proving the conjectured formula for the case $q=m$ and verifying that the observed behavior as $q$ increases holds in all dimensions. 

On more general \kahler manifolds the formula for the expected number $\ncal^\crit_{N,h}$ of critical points is much more difficult to evaluate.  Because things simplify as the degree of the bundle gets large, an asymptotic expansion of $\ncal^\crit_{N,h}$ and an integral formula for its leading coefficient $b_0$ were derived in \cite{DSZ2}.  The leading coefficient was shown to be universal and therefore, based on calculations on $\CP^m$, it was conjectured that the critical points of minimal Morse index were the most plentiful as $N \to \infty$, and upper and lower asymptotic bounds for $\ncal^\crit_{N,h}$ were estimated in Conjecture 4.4 of \cite{DSZ2}.  We are able to apply our methods to $b_0$ and work out a proof of this conjecture as well, with an improvement on the upper bound estimate.

\subsection{Background}\label{A2}

The setting for this paper is the $N$th tensor power of a positive Hermitian line bundle $(L^N,h^N) \rightarrow (M^m,\omega_h)$ over a compact \kahler manifold of dimension $m$.  Here $\omega_h$ is the \kahler form and is given by $\omega_h = \frac{i}{2} \Th_{h}$, where $\Th_{h} = - \partial \bar \partial \log h$ is the curvature form of the metric.  The connection on the bundle is taken to be the Chern connection $\nabla$ of $h^N$.  Relative to this connection, the critical points of a holomorphic section $s \in H^0(M,L^N)$ are given by $\nabla s(z)=0$ and the set of critical points of $s$ will be denoted by $Crit(s,h^N)$. 

We note that in general, the critical point equation is not holomorphic and thus the cardinality of $Crit(s,h^N)$ is a non-constant random variable on the space $H^0(M,L^N)$.  Indeed we see that in a local frame $e$ we can write $s=fe$ and then
$ \nabla s = \left( \partial f - f \partial K \right) \otimes e_L$, where $K = - \log \| e \|_{h^N}^2$ is the \kahler potential.
Thus the critical point equation in the local frame is $\partial f - f \partial K = 0$, which is holomorphic only when $K$ is.

The space $H^0(M,L^N)$ is endowed with the Gaussian measure $\gamma_{N}$ given by \begin{equation*}d\gamma_{N}(s)=\frac 1
{\pi^d}e^{-\|c\|^2} dc\;,\qquad  s=\sum _{j=1}^d
c_je_j,\end{equation*} where $dc$ is Lebesgue measure and $\{e_j\}$
is an orthonormal basis of $H^0(M,L^N)$ relative to the inner product \begin{equation*} \langle s_1, s_2 \rangle = \frac{1}{m!} \int_M h^N(s_1(z), s_2(z)) \ \omega_h^m \end{equation*} induced by $h^N$ on $H^0(M,L^N)$. 
The expected distribution of critical points of $s \in
H^0(M, L^N)$ with respect to  $\gamma_{N}$ is defined to be
\begin{equation*}  \K^\crit_{N,h}  =\int_{H^0(M, L)}
 \bigg[\,\sum_{z\in Crit (s, \, h^N)} \delta_{z}\bigg]\,d\ga_N(s),
\end{equation*} 
where $\delta_{z}$ is the Dirac point mass at $z$, and the expected total number of critical points is then given by \begin{equation*}\ncal^\crit_{N,h}=\int_M \K^\crit_{N,h}(z).
\end{equation*}

The critical points of $s$ with respect to $\nabla$ are the same as those of $\log \|s\|_{h^N}^2$, and therefore as an aid in the analysis of the statistics of the critical points we consider their Morse indices.  Recall that the Morse index $q$ of a critical point of a real-valued function is given by the number of negative eigenvalues of its Hessian.  For a positive line bundle it is well-known that $m \leq q \leq 2m$ \cite{Bo}.  We let $\K^\crit_{N,q,h}$ denote the expected distribution of critical points of Morse index $q$, and $\ncal^\crit_{N,q,h}$ denote the expected number of these critical points.  It follows that
\begin{equation*} \K^\crit_{N,h}(z)=\sum_{q=m}^{2m}\K^\crit_{N,q,h}(z)\,,\qquad
\ncal^\crit_{N,h}=\sum_{q=m}^{2m}\ncal^\crit_{N,q,h}\;.\end{equation*}

We now recall the relevant results from \cite{DSZ2}.  First, we have the integral formula for $\ncal^\crit_{N,q,h}(\CP^m)$.

\begin{theo} \label{DSZTheo3}   The expected number of critical points  of Morse
index $q$ for random sections $s\in H^0(\CP^m, \ocal(N))$ is
given by   \begin{eqnarray*}
{\mathcal  N}^\crit_{N,q,h}(\CP^m) &=& \frac{ 2^{\frac
{m^2+m+2}2}}{\prod_{j=1}^m j!}\ \frac{(N-1)^{m+1}
}{(m+2)N-2}
\int_{Y_{2m-q} }d\la\,
\left|{\textstyle \prod_{j=1}^m\lambda_j} \right| \,\De(\lambda)\,e^{
-\sum_{j=1}^m \la_j} \\ && \phantom{\frac{ 2^{\frac
{m^2+m+2}2}}{\prod_{j=1}^m j!}\ \frac{(N-1)^{m+1}
}{(m+2)N-2}
\int_{Y_{2m-q} }} \times
\begin{cases} e^{(m+2-2/N) \lambda_m}
& \!\!\text{for $q>m$}
\\ 1 & \!\!\text{for $q=m$} \end{cases}
\end{eqnarray*} for $N\ge 2$, where
$Y_p=\{\la\in\R^m: \la_1>\cdots >\la_p>0>\la_{p+1}>\cdots
>\la_m\}$ and
$\De(\lambda) =
\prod_{i < j} (\lambda_i - \lambda_j)$ is the Vandermonde
determinant.
\end{theo}

Next we have the complete asymptotic expansions  of $\K_{N,q,h}^\crit(z)$ and $\ncal^\crit_{N,q,h}$ on a general \kahler manifold.

\begin{theo} \label{DSZTheo1}  For any positive Hermitian  
line bundle $(L, h) \to (M,
\omega_h)$ over any compact \kahler manifold with $\om_h = \frac
i2\Th_h$, the expected distribution of critical points of
Morse index $q$ of random sections in $H^0(M,L^N)$ relative to the Hermitian Gaussian measure induced by $h$ and $\om_h$ has an asymptotic expansion of the form 
$$N^{-m}\,\K_{N,q,h}^\crit(z)\sim \{b_{0q}  + b_{1q}(z) N\inv  + b_{2q}(z)
N^{-2} + \cdots\}\frac {\om_h^m}{m!}\;,\qquad m\le q\le 2m\;,$$ where
the $b_{jq}=b_{jq}(m)$ are curvature invariants of order $j$ of
$\omega_h$. In
particular, $b_{0q}$ is the universal constant
\begin{equation}\label{b0q}b_{0q} = \pi^{-\binom{m+2}{2}}  \int_{{\bf S}_{m,q-m}} \left|\det(2H H {}^*-|x|^2I)\right|\,e^{- \langle (H,
x), (H,x) \rangle} \,dH\,dx\,,\end{equation}
where \begin{equation*}{\bf S}_{m,k}:=\{(H,x)\in
\sym(m,\C)\times\C: \mbox{\rm index}(2HH^*-|x|^2I)
=k\}\;.\end{equation*}
\end{theo}

\begin{cor} \label{DSZTheo2}  Let $(L,h)\to (M,\om_h)$ be a
positive holomorphic line bundle on a compact \kahler manifold,
with $\om_h= \frac i2 \Th_h$. Then the expected  number of
critical points of Morse index $q$ ($m\le q\le 2m$) of random
sections in $H^0(M,L^N)$ has the asymptotic expansion
\begin{eqnarray}\ncal^\crit_{N,q,h} &\sim
&\left[\frac{\pi^m b_{0q}}{m!}\,c_1(L)^m\right]N^m +
\left[\frac{\pi^m\be_{1q}}{(m-1)!}\, c_1(M)\cdot
c_1(L)^{m-1}\right]N^{m-1} \nonumber
\\& & \ + \biggl[\be_{2q}\int_M\rho^2d\vol_h + \be'_{2q}\, c_1(M)^2\cdot
c_1(L)^{m-2} \nonumber \\ & & \qquad \qquad \qquad \qquad \qquad \qquad + \be''_{2q}\, c_2(M)\cdot c_1(L)^{m-2}\biggr]
N^{m-2}+\cdots\,,\nonumber\end{eqnarray} where
$b_{0q},\be_{1q},\be_{2q},\be'_{2q},\be''_{2q}$ are universal constants
depending only on the dimension $m$.
\end{cor}

We ask the interested reader to refer to \cite{DSZ1,DSZ2,DSZ3} for additional background information.

\subsection{Results}\label{A3}

Our first result gives the exact formula for  $\ncal^\crit_{N,q,h}(\CP^m)$ when $q=m$, shows that this number decreases as $q$ increases, and gives an upper and lower bound for the total expected number $\ncal^\crit_{N,h}(\CP^m)$ which holds for all $N$ and $m$.

\begin{theo} \label{CPm}   Let ${\mathcal  N}^\crit_{N,q,h}(\CP^m)$ denote the expected number of critical points of Morse
index $q$ for random sections $s\in H^0(\CP^m, \ocal(N)),$ then
\begin{equation*} 
{\mathcal  N}^\crit_{N,m,h}(\CP^m) = \frac{2(m+1)(N-1)^{m+1}}{(m+2)N-2},  
\end{equation*}
and when $N>2$ 
\begin{equation*} 
{\mathcal  N}^\crit_{N,q+1,h}(\CP^m)  <  {\mathcal  N}^\crit_{N,q,h}(\CP^m).
\end{equation*}
Therefore, 
\begin{equation*} 
\frac{2(m+1)(N-1)^{m+1}}{(m+2)N-2} < {\mathcal  N}^\crit_{N,h}(\CP^m)  < \frac{2(m+1)^2 (N-1)^{m+1}}{(m+2)N-2}.
\end{equation*}
\end{theo}

In order to obtain the exact formula for ${\mathcal  N}^\crit_{N,m,h}(\CP^m)$ we apply a modification of Selberg's integral formula to the integral in Theorem \ref{DSZTheo3}.  The second part of the theorem then follows from a change of variable argument.  These arguments are presented in \S \ref{B1}.

From this theorem we see that ${\mathcal  N}^\crit_{N,q,h}(\CP^m)$ grows exponentially with the dimension.  This verifies the conjectured growth rate for ${\mathcal  N}^\crit_{N,q,h}(\CP^m)$ that was used in \cite[Sec. 7.3]{DSZ3} as a basis for their heuristic estimate of the growth rate for the expected density of vacua in string/M theory.    

The modulus of the spectral determinant shows up in the various integral formulas for the expected number of critical points (\cite{AD}, \cite {BM}, \cite{DSZ2}, \cite{F}).  As the modulus presents a serious technical challenge in evaluating the integral, it is often dropped from the calculation (see \cite{AD} and \cite {BM}), which results in counting the critical points with signs.  In string theory this is known as computing the ``supergravity index'', while in spin glass theory there is some debate over the validity and implications of the calculation (see \cite{ABM} and references therein).    
In our case, Morse theory tells us that the number of critical points of each $s\in H^0(\CP^m, \ocal(N))$ counted with signs is a topological invariant and is given by 
\begin{multline*}
\sum_{z: \nabla s(z) = 0} (-1)^q = c_m(T^{*1,0}_{\CP^m}\otimes \ocal(N)) \\ = \sum_{j=0}^m (-1)^j \left(\genfrac{}{}{0cm}{1}{m+1}{j} \right) N^{m-j} = \frac{(N-1)^{m+1} + (-1)^m}{N},
\end{multline*}
where $q$ is the Morse index of $z$.  We see that this ``index counting'' provides a good estimate of the total expected number of critical points, giving the correct growth rate except for the coefficient.

Next we turn our attention to $b_{0q}$ and note that the absolute value sign in \eqref{b0q} prevents the direct application of Wick methods.  Therefore in \S \ref{C1} we utilize a variant of the Itzykson-Zuber formula in random matrix integrals, as was done in the simplification of the formula for $\beta_{2q}$ in \cite{DSZ2}, to derive the following alternative formula for $b_{0q}$. 

\begin{theo}\label{b0qtheo}  In all dimensions,
\begin{align*}
b_{0q}(m) &= \frac{2^{\frac {m^2+m+2}2}}{\pi^m(m+2)\prod_{j=1}^{m-1} j!}
\\ & \quad \times \int_{Y_{2m-q} }d\la\,
\left|{\textstyle \prod_{j=1}^m\lambda_j} \right| \,\De(\lambda)\,e^{
-\sum_{j=1}^m \la_j} \times
\begin{cases} e^{(m+2) \lambda_m}
& \!\!\text{for $q>m$}
\\ 1 & \!\!\text{for $q=m$} \end{cases}.
\end{align*}
Here
$Y_p=\{\la\in\R^m: \la_1>\cdots >\la_p>0>\la_{p+1}>\cdots
>\la_m\}\;$
and
$\De(\lambda) =
\prod_{i < j} (\lambda_i - \lambda_j)$ is the Vandermonde
determinant.
\end{theo}

We see that the integral in the above theorem is almost identical to the one in Theorem \ref{DSZTheo3}, so we apply the methods of \S \ref{B1} to this formula to obtain:

\begin{theo} \label{Conj}
Let $n_q(m):=\frac{\pi^m}{m!}\,b_{0q}(m)$ denote the
leading coefficient in the expansion of $\ncal^\crit_{N,q,h}$, and let $n(m)=\sum_{q=m}^{2m} n_q(m)$: $$\ncal^\crit_{N,q,h}
\sim n_q(m)\,c_1(L)^m\,N^m ,\qquad \ncal^\crit_{N,h}
\sim n(m)\,c_1(L)^m\,N^m .$$  Then $$n_m(m)=
2\,\frac{m+1}{m+2} \qquad \text{and} \qquad n_{q+1}(m) <
 \left(\frac{2m-q}{2m-q+1}\right)^2 n_{q}(m)\;,$$ and hence the expected total number of critical
points
$$\ncal^\crit_{N,h} \sim
n(m)\,c_1(L)^m\,N^m \qquad \text{with} \qquad 2\,\frac{m+1}{m+2} <n(m) <
\frac{2m+3}{3}\;.$$
\end{theo}

This theorem proves Conjecture 4.4 in \cite{DSZ2} which was made based on calculations in small dimensions for the leading coefficient in the $\CP^m$ case.

\smallskip

These results are part of the author's ongoing thesis research at the Johns Hopkins University which is being advised by S. Zelditch.

\section{Proof of Theorem \ref{CPm}}\label{B1}

In this proof we first work out the formula for the minimal Morse index case ${\mathcal  N}^\crit_{N,m,h}(\CP^m)$ and then proceed to show that ${\mathcal  N}^\crit_{N,q,h}(\CP^m) > {\mathcal  N}^\crit_{N,q+1,h}(\CP^m)$ for $m \leq q \leq 2m$.  The proofs of the intermediate lemmas will be given in the subsections below. 

From Theorem \ref{DSZTheo3} we have
\begin{equation*} 
{\mathcal  N}^\crit_{N,m,h}(\CP^m) = \frac{ 2^{\frac
{m^2+m+2}2}}{\prod_{j=1}^m j!}\ \frac{(N-1)^{m+1}
}{(m+2)N-2} \int_{Y_{m} }
\left|{\textstyle \prod_{j=1}^m\lambda_j} \right| \,\De(\lambda)\,e^{
-\sum_{j=1}^m \la_j} d\la. 
\end{equation*}
We then use 
\begin{lem} \label{q=m} 
\begin{equation} \label{q=mformula}
\frac{ 2^{\frac{m^2+m+2}2}}{\prod_{j=1}^m j!}\
\int_{0 < \la_m < \dots < \la_1 < \infty}
{\textstyle \prod_{j=1}^m \la_j}  \  \De(\la)\,e^{
-\sum_{j=1}^m \la_j} \ d\la = 2(m+1)
\end{equation}
\end{lem}
\noindent to obtain 
\begin{equation} \label{CPmm}
{\mathcal  N}^\crit_{N,m,h}(\CP^m) = \frac{2(m+1)(N-1)^{m+1}}{(m+2)N-2} \, .  
\end{equation}

For the general case, we recall that
\begin{eqnarray}
{\mathcal  N}^\crit_{N,q,h}(\CP^m) &=& \frac{ 2^{\frac
{m^2+m+2}2}}{\prod_{j=1}^m j!}\ \frac{(N-1)^{m+1}
}{(m+2)N-2}
\int_{Y_{2m-q}} \!\!\!\! d\la\,
\left|{\textstyle \prod_{j=1}^m\lambda_j} \right| \,\De(\lambda)\,e^{
-\sum_{j=1}^m \la_j} \label{CPmformula}
\\ && \phantom{\frac{ 2^{\frac
{m^2+m+2}2}}{\prod_{j=1}^m j!}\ \frac{(N-1)^{m+1}
}{(m+2)N-2}
\int_{Y_{2m-q} }} \times
\begin{cases} e^{(m+2-2/N) \lambda_m}
& \!\!\text{for $q>m$}
\\ 1 & \!\!\text{for $q=m$} \end{cases} \nonumber .
\end{eqnarray}
We will show in \S \ref{B3} that:
\begin{lem} \label{q>q+1} 
For $m \geq 1$ and $0 \leq p \leq m$, let
\begin{equation*} 
P_{c,\,p}(m) = \int_{Y_{p} } \!\!\! d\la\,
\left| \prod_{j=1}^m\lambda_j \right| \,\De(\lambda)\,e^{
-\sum_{j=1}^{m-1} \la_j} \times
\begin{cases} e^{(m+c) \lambda_m} & \text{for $p<m$}
\\e^{-\lambda_m} & \text{for $p=m$}
\end{cases} \, ,
\end{equation*}
where $Y_p$ is as in Theorem \ref{DSZTheo3}.  Then $P_{0,\,r}(m) = P_{0,\,s}(m)$ for $0 \leq r,s \leq m$, and for $c > 0$
\begin{equation*} 
P_{c,\,p-1}(m) < \left(\frac{p}{p+c}\right)^2 P_{c,\,p}(m).
\end{equation*}
\end{lem}
From this we see first that for $N=2$ we can apply the above lemma with $p=2m-q$ and $c=0$ to the integral in \eqref{CPmformula}.  Thus ${\mathcal  N}^\crit_{2,r,h}(\CP^m) = {\mathcal  N}^\crit_{2,s,h}(\CP^m)$ for $m \leq r,s \leq 2m$.  From \eqref{CPmm} we see that 
${\mathcal  N}^\crit_{2,m,h}(\CP^m) = 1$ and therefore $${\mathcal  N}^\crit_{2,h} (\CP^m) \equiv \sum_{q=m}^{2m} {\mathcal  N}^\crit_{2,q,h} (\CP^m) = m+1$$ for $m \geq 1$.  

Then, when $N>2$, we apply Lemma \ref{q>q+1} with $p=2m-q$ and $c=1 - \frac{2}{N} \,$ to obtain
\begin{equation*} 
{\mathcal  N}^\crit_{N,q+1,h}(\CP^m)  <  \left( \frac{2m-q}{2m-q + 1 - \frac{2}{N}} \right)^2 {\mathcal  N}^\crit_{N,q,h}(\CP^m).
\end{equation*}
Therefore,
\begin{equation*} 
\frac{2(m+1)(N-1)^{m+1}}{(m+2)N-2} < {\mathcal  N}^\crit_{N,h}(\CP^m)  < \frac{2(m+1)^2 (N-1)^{m+1}}{(m+2)N-2} \, .
\end{equation*}

\subsection{Proof of Lemma \ref{q=m}}\label{B2}

First, we need the following well-known theorem (see \cite{Se}).
\begin{theo} [\textbf{Selberg's Integral Formula}]\label{Selberg}
For any positive integer n, let 
\begin{equation*}  
\Ph(\la) \equiv \Ph(\la_1, \cdots , \la_n) = \left|  \De(\la)   \right|^{2\ga} \prod_{j=1}^n \la_j^{\al-1} (1-\la_j)^{\be - 1}.
\end{equation*}
Then 
\begin{equation}  \label{Selbergformula} 
\int_0^1 \cdots \int_0^1 \Ph(\la) d\la = \prod _{j=0}^{n-1} \frac{\Ga(1+\ga+j\ga)\Ga(\al+j\ga)\Ga(\be+j\ga)} {\Ga(1+\ga)\Ga(\al+\be+\ga(n+j-1))}, 
\end{equation}
when $\al, \be, \ga \in \C$ with Re $\al >0$, Re $\be >0$, Re $\ga >$ -min $\left( \frac{1}{n}, \frac{Re \al}{(n-1)},  \frac{Re \be}{(n-1)} \right)$.
\end{theo}
\noindent As a corollary, we have a special limiting case of the above formula (see \cite{As}).

\begin{cor}\label{SelbergCor}
For any positive integer n, let 
\begin{equation*}  
\Ph(\la) \equiv \Ph(\la_1, \cdots , \la_n) = \left|\De(\la)\right|^{2\ga} \prod_{j=1}^n \la_j^{\al-1} \, e^{-\la_j} .
\end{equation*}
Then 
\begin{equation}  \label{Selbergexpformula} 
\int_0^\infty \cdots \int_0^\infty \Ph(\la) d\la = \prod _{j=0}^{n-1} \frac{\Ga(1+\ga+j\ga)\Ga(\al+j\ga)} {\Ga(1+\ga)}, 
\end{equation}
valid for complex $\al$, $\ga$ with Re $\al >0$, Re $\ga >$ -min $\left( \frac{1}{n}, \frac{Re \, \al}{(n-1)} \right)$.
\end{cor}

\noindent This formula is obtained by setting $\be = m$ and making the change of variables $x_j \to \frac {x_j}{m}$ in \eqref{Selbergformula} and then letting $m \to \infty$.
 
In order to simplify the notation we will use $P(m)$ to denote the integral on the LHS of \eqref{q=mformula}.  We see that we can rewrite this integral as
\begin{equation*} 
P(m) = \int_{0 < \la_m < \dots < \la_1 < \infty}
{\textstyle \prod_{j=1}^m\lambda_j} \,\left| \De(\lambda) \right| \,e^{-\sum_{j=1}^m \la_j}d\la.
\end{equation*}
We then note that the integrand on the RHS of the above equation is symmetric under permutations of $\la$.  Therefore,
\begin{equation} \label{Pmformula}
P(m) = \frac{1}{m!} \int_{\R_+^m}
{\textstyle \prod_{j=1}^m\lambda_j} \,\left| \De(\lambda) \right| \,e^{-\sum_{j=1}^m \la_j}d\la.
\end{equation}
It is easy to see that the integrals in \eqref{Selbergexpformula} and \eqref{Pmformula} are equal when $\al=2$, $\ga=\frac{1}{2}$, and $n=m$.  
Consequently,
\begin{equation*}  
P(m) = \frac{1}{m!}\prod_{j=0}^{m-1} \frac{\Ga(\frac{3}{2}+\frac{j}{2})\Ga(2+\frac{j}{2})}{\Ga(\frac{3}{2})}=(m+1)\prod_{j=1}^{m}2^{-j}j!\,,
\end{equation*} 
where the last equality follows from an application of Gauss's multiplication formula.  The desired formula is then obtained by substituting $P(m)$ back into \eqref{q=mformula}.

\subsection{Proof of Lemma \ref{q>q+1} } \label{B3} 

In order to simplify the discussion we will examine the case where $p=m$ separately from the others.  In this case
\begin{equation*}
P_{c,\, m}(m) = \int_{0<\la_m<\dots<\la_1<\infty } \,
\left(\prod_{j=1}^m\lambda_j\right) \left( \prod_{i=1}^{m-1} \prod_{j=i+1}^m (\lambda_i - \lambda_j) \right)\,e^{
-\sum_{j=1}^m \la_j} d\la .
\end{equation*}
We make the change of variables $$ \left\{ \la_1, \dots , \la_m \right\} \rightarrow \left\{ \sum_{i=1}^m \la_i , \sum_{i=2}^m \la_i , \dots , \la_m \right\}$$ to obtain
\begin{equation*}
P_{c,\, m}(m) = \int_{\R_+^m } \,
\left( \prod_{i=1}^m \sum_{j=i}^m \lambda_j \right) 
\left( \prod_{i=1}^{m-1} \prod_{j=i}^{m-1} \sum_{k=i}^j \la_k \right)\,e^{-\sum_{j=1}^m j\,\la_j} d\la \, .
\end{equation*}
Next we see that 
\begin{align}
\left( \prod_{i=1}^m \sum_{j=i}^m \lambda_j \right) 
\left( \prod_{i=1}^{m-1} \prod_{j=i}^{m-1} \sum_{k=i}^j \la_k \right) & = \la_m \left( \prod_{i=1}^{m-1} \sum_{j=i}^m \lambda_j \right) \left( \prod_{i=1}^{m-1} \prod_{j=i}^{m-1} \sum_{k=i}^j \la_k \right) \nonumber \\ & = \la_m \left( \prod_{i=1}^{m-1} \prod_{j=i}^{m} \sum_{k=i}^j \la_k \right) \label{prodsimp} \\ & = \prod_{i=1}^{m} \prod_{j=i}^{m} \sum_{k=i}^j \la_k \nonumber
\end{align}
and so
\begin{equation}\label{p=m}
P_{c,\, m}(m) = \int_{\R_+^m } \,
\left( \prod_{i=1}^{m} \prod_{j=i}^{m} \sum_{k=i}^j \la_k \right)\,e^{-\sum_{j=1}^m j\,\la_j} d\la \, .
\end{equation}

Now we consider $P_{c,\, p}(m)$ when $0 \leq p<m$.  For these cases
\begin{equation*}
P_{c,\, p}(m) = \int_{Y_p } \,
\left| \prod_{j=1}^m\lambda_j \right| \left( \prod_{i=1}^{m-1} \prod_{j=i+1}^m (\lambda_i - \lambda_j) \right) e^{
-\sum_{j=1}^{m-1} \la_j + (m+c) \la_m} d\la ,
\end{equation*}
and we make the change of variables
$$ \left\{ \la_1, \dots , \la_p \right\} \rightarrow \left\{ \sum_{i=1}^p \la_i , \sum_{i=2}^p \la_i , \dots , \la_p \right\}, $$ $$ \left\{ \la_{p+1}, \dots , \la_m \right\} \rightarrow \left\{ -\la_{p+1}, -(\la_{p+1}+\la_{p+2}), \dots , -\sum_{i=p+1}^m \la_i  \right\} $$
to obtain
\begin{multline*}
P_{c,\, p}(m) = \int_{\R_+^m } \,
\left( \prod_{i=1}^p \sum_{j=i}^p \lambda_j \right) 
\left( \prod_{i=p+1}^m \sum_{j=p+1}^i \lambda_j \right) 
\left(\prod_{i=1}^{p-1} \prod_{j=i}^{p-1} \sum_{k=i}^j \la_k \right)
\\ \times
\left( \prod_{i=1}^{p} \prod_{j=p+1}^{m} \sum_{k=i}^j \la_k \right)
\left( \prod_{i=p+2}^{m} \prod_{j=i}^{m} \sum_{k=i}^j \la_k \right)
\,e^{-\sum_{j=1}^p j\,\la_j - \sum_{j=p+1}^m (j+c)\,\la_j} d\la \, .
\end{multline*}
We can combine the first quantity with the third, and the second with the fifth, as we did in \eqref{prodsimp}, and thus
\begin{multline*}
P_{c,\, p}(m) = \int_{\R_+^m } \!
\left(\prod_{i=1}^{p} \prod_{j=i}^{p} \sum_{k=i}^j \la_k \right)\!\!
\left( \prod_{i=p+1}^{m} \prod_{j=i}^{m} \sum_{k=i}^j \la_k \right)\!\!
\\ \times
\left( \prod_{i=1}^{p} \prod_{j=p+1}^{m} \sum_{k=i}^j \la_k \right)
\! e^{-\sum_{j=1}^p j\,\la_j - \sum_{j=p+1}^m (j+c)\,\la_j} d\la \, .
\end{multline*}
Now it is clear that
\begin{equation*}
\left(\prod_{i=1}^{p} \prod_{j=i}^{p} \sum_{k=i}^j \la_k \right)
\left( \prod_{i=1}^{p} \prod_{j=p+1}^{m} \sum_{k=i}^j \la_k \right)
= \prod_{i=1}^{p} \prod_{j=i}^{m} \sum_{k=i}^j \la_k \, ,
\end{equation*}
and then 
\begin{equation*}
\left( \prod_{i=1}^{p} \prod_{j=i}^{m} \sum_{k=i}^j \la_k \right)
\left( \prod_{i=p+1}^{m} \prod_{j=i}^{m} \sum_{k=i}^j \la_k \right)
= \prod_{i=1}^{m} \prod_{j=i}^{m} \sum_{k=i}^j \la_k \, .
\end{equation*}
Therefore,
\begin{equation} \label{P_presult}
P_{c,\, p}(m) = \int_{\R_+^m } \!
\left(\prod_{i=1}^{m} \prod_{j=i}^{m} \sum_{k=i}^j \la_k \right) e^{-\sum_{j=1}^p j\,\la_j - \sum_{j=p+1}^m (j+c)\,\la_j} d\la .
\end{equation}
We note that in this formula the only dependence on $p$ is in the exponential, and we see from  \eqref{p=m} that this formula also holds for the $p=m$ case as well.  

When $c=0$, the formula does not depend on $p$ at all, so we see that $P_{0,\, r}(m) = P_{0,\, s}(m)$ for $0 \leq r,s \leq m$.

Next we let $c>0$ and rewrite \eqref{P_presult} as follows, 
\begin{equation*} 
P_{c,\, p}(m) = \int_{\R_+^m } \! \ical(\la_1, \ldots, \la_m)
\left(\prod_{i=1}^{m} \la_i \right)  e^{-\sum_{j=1}^{p} j\,\la_j - \sum_{j=p+1}^m (j+c)\,\la_j} d\la \, ,
\end{equation*}
where
\begin{equation*} 
\ical(\la_1, \ldots, \la_m) = \prod_{i=1}^{m} \prod_{j=i+1}^{m} \sum_{k=i}^j \la_k .
\end{equation*}
Then we make the change of variable $\la_{p} \rightarrow \frac{p}{p+c} \la_{p}$ in the formula for $P_{c, \, p-1}$ to obtain
\begin{align*} 
P_{c,\, p-1}(m) &= \left(\frac{p}{p+c}\right)^2 
\int_{\R_+^m } \! \ical(\la_1, \ldots, {\textstyle \frac{p}{p+c}}\la_{p}, \ldots, \la_m)
\\ & \quad \phantom{\left(\frac{p}{p+c}\right)^2 
\int_{\R_+^m }} \times \left(\prod_{i=1}^{m} \la_i \right)  e^{-\sum_{i=1}^{p} i\,\la_i - \sum_{j=p+1}^m (j+c)\,\la_j} d\la 
\\ & < \left(\frac{p}{p+c}\right)^2 P_{c,\, p}(m).
\end{align*}

\section{Proof of Theorem \ref{b0qtheo}}\label{C1}

We begin with an intermediate lemma.  This formula follows with only slight modifications from the derivation given in \cite[Sec. 6.3]{DSZ2} of a similar formula for the constant $\be_{2q}$.  For the sake of completeness we present the entire proof below.

\begin{lem} \label{b0qlemma}
\begin{align*} b_{0q}(m) &= \frac {(-i)^{m(m-1)/2}} {\pi^{2m}\prod_{j=1}^{m-1} j!}
\\ & \quad \times \int_{Y_{2m-q}}\int_\R \!\!\cdots\!\int_\R \frac{\De(\la)\,\De(\xi)\,
\prod_{j=1}^m |\la_j|\;e^{i\langle
\la,\xi\rangle}}{\left(1 - \frac i2 \sum_j\xi_j\right)\prod_{j\le k}\left[ 1
+\frac i2 (\xi_j+\xi_k)\right] }\, \,d\xi_1\cdots d\xi_m\,
d\la\;.\end{align*} 
Here, $\De(\lambda)$ and $Y_{2m-q}$ are as in Theorem \ref{b0qtheo}, and the iterated $d\xi_j$ integrals are defined in the
distribution sense.
\end{lem}

\begin{proof}

First, we let \begin{multline}
\ical_{\ep,\ep'}  =  \frac 1{\pi^{d_m}} \int_{{\mathcal H}_m}
\int_{\hcal_m(m-q)}\int_{\sym(m,\C)\times\C} \textstyle
\: \left|\det(2P)\right|
e^{i \langle \X, P - HH^* +\half |x|^2 I\rangle}  \\ \times \ e^{ -\tr  HH^*-|x|^2}\, e^{- \epsilon \tr \,
\X\X^*-\ep' \tr  PP^* }\, dH \,dx \,dP \,d
\X\;, \end{multline}
where ${\mathcal H}_m$ is the space of $m \times m$
Hermitian matrices,
$\hcal_m(m-q)= \{P\in
\hcal_m:\mbox{index}\,P=m-q\},$ and $ d_m = \dim_\C(\sym(m,\C)\times\C) = \half(m^2+m+2).$
We note that absolute convergence in the above integral is guaranteed
by the Gaussian factors in each of the variables $(H, x, P,\X)$.
Then we have that 
\begin{equation}\label{beta-ical} b_{0q}(m) = 
\frac 1{\pi^m\,(2\pi)^{m^2}}\; \lim_{\ep'\to 0}\;\lim_{\ep\to
0}\ical_{\ep,\ep'}\;.\end{equation}
To verify this, first evaluate $\int e^{i \langle \X, P - HH^* +\half
|x|^2\rangle}e^{-\ep\tr \, \X\X^*} d\X$ to obtain a dual
Gaussian, which  approximates the delta function $\de_{HH^* -\half
|x|^2}(P)$. As $\epsilon \to 0$, the $d P$ integral then yields
the integrand at $P = H H^*-\half|x|^2I$; then we let $\ep'\to 0$
to obtain the original integral.

Next we  conjugate $P$ to a diagonal
matrix $D(\lambda)$ with $\lambda = (\lambda_1, \dots, \lambda_m)$
by an element $h \in \U(m)$. We recall that
\begin{equation} \label{IDENTITY} \int_{\hcal_m}\phi(P)\,dP =
c_m' \int_{\R^m}\int_{\U(m)}\phi(
hD(\lambda)h^*)\De(\lambda)^2\,dh\,d\la\,, \quad c_m'= \frac{(2\pi)^{\binom{m}{ 2}}}
{\prod_{j=1}^m j!},  \end{equation}
where $dh$ is unit mass Haar measure on $\U(m)$ (see for example \cite[(1.9)]{ZZ}), and use this to obtain
\begin{eqnarray*}
\ical_{\ep,\ep'} & = &  \frac{2^m\,c_m'}{\pi^{d_m}} \int_{\U(m)}
\int_{{\mathcal H}_m}
\int_{Y'_{2m-q}}\int_{\sym(m,\C)\times\C}\De(\la)^2\,
\prod_{j=1}^m |\la_j|\,\textstyle \, e^{-\tr  HH^*-|x|^2} \\
&& \qquad \times\ e^{i\langle \X,\, hD(\la)h^* +\half |x|^2I- HH^*
\rangle} e^{ -\left[\epsilon \tr \X\X^*
+\epsilon'  \sum \la_j^2\right]}\, dH \,dx \,d\la \,d \X\,dh\,.
\end{eqnarray*}
Here $Y'_p$ denotes the set of points in $\R^m$ with
exactly $p$ coordinates positive.  Again using \eqref{IDENTITY} applied this time to $\X$, we
obtain
\begin{eqnarray*}
\ical_{\ep,\ep'}& = &  \frac{ 2^m(c'_m)^2}{\pi^{d_m}} \int_{\U(m)}
\int_{\U(m)} \int_{\R^{m}}\int_{Y'_{2m-q}}
\int_{\sym(m,\C)\times\C}\De(\la)^2\,\De(\xi)^2\,
\prod_{j=1}^m |\la_j|\,\textstyle \:
\\&& \qquad \qquad  \qquad\times\
e^{i \langle gD(\xi)g^*,\, hD(\la)h^* +\half |x|^2I- HH^*\rangle}
\\&& \qquad \qquad \qquad \times\
e^{ -\tr  HH^*-|x|^2-  \sum(\ep\xi_j^2 +\ep'\la_j^2)} \, dH \,dx
\,d\la\,d \xi\,dh\,dg\;.\end{eqnarray*}

 We then  transfer the conjugation  by $g$ to the right side of the
$\langle, \rangle$ in the first exponent and make the change of
variables $h\mapsto gh, H\mapsto gHg^t$ to eliminate $g$ from the
integrand:
\begin{eqnarray*} \ical_{\ep,\ep'}& = &   \frac{2^m(c'_m)^2 }{\pi^{d_m}}
\int_{\U(m)} \int_{\R^{m}}\int_{Y'_{2m-q}}
\int_{\sym(m,\C)\times\C}\De(\la)^2\,\De(\xi)^2\,
\prod_{j=1}^m |\la_j|\;\textstyle \:
\\&&
\times\ e^{i \langle D(\xi),\, hD(\la)h^* +\half |x|^2I-
HH^*\rangle} e^{ -\tr  HH^*-|x|^2- \sum(\ep\xi_j^2 +\ep'\la_j^2)}
\, dH \,dx \,d\la\,d \xi\,dh\;.
\end{eqnarray*}

Next we recognize the integral  $\int_{\U(m)} e^{i \langle D(\xi),
h D(\lambda) h^* \rangle}dh$ as   the well-known
Itzykson-Zuber-Harish-Chandra integral \cite{Ha} (cf., \cite{ZZ}):
\begin{equation}\label{IZ}J(D(\lambda), D(\xi))   =
(-i)^{m(m-1)/2}\left({\textstyle \prod_{j=1}^{m-1}j!}\right)\frac{\det
[e^{i\lambda_j
\xi_k}]_{j,k}}{\De(\lambda) \De(\xi)}\;.\end{equation}
We substitute \eqref{IZ} into the above integral and expand
$$\det [e^{i
\xi_j \lambda_k}]_{jk} = \sum_{\sigma \in S_m} (-1)^\sigma\;
e^{i\langle  \xi, \sigma(\lambda)\rangle }, $$  obtaining a sum of
$m!$ integrals.  However, by making the change of variables
$\sigma(\la) \to \la'$ and noting that $\De(\la')=  (-1)^\sigma
\De(\la)$, we see that the integrals of all these terms are
equal, and so we obtain
\begin{eqnarray*}
\ical_{\ep,\ep'}& = &    (-i)^{m(m-1)/2}\frac{c_m''}{\pi^{d_m}}
\int_{\R^{m}}\int_{Y'_{2m-q}}
\int_{\sym(m,\C)\times\C}\De(\la)\,\De(\xi)\, \prod_{j=1}^m
|\la_j|\;e^{i\langle
\la,\xi\rangle}\\
&& \textstyle\times\ \: \exp\left(i
\left\langle D(\xi),\,\half |x|^2I-HH^* \right\rangle -\tr
HH^*-|x|^2\right)\\&& \textstyle\times\ \exp\left( -\ep
\sum\xi_j^2 -\ep'\sum\la_j^2\right) \, dH \,dx \,d\la\,d
\xi\,,
\end{eqnarray*} where
$$c_m''= \frac {2^{m^2}\,\pi^{m(m-1)}}
{\prod_{j=1}^m j!}\;.$$

The phase 
\begin{eqnarray*}\Ph (H,x;\xi) &:=& i \left\langle D(\xi),
 \half |x|^2 I  -  HH^* \right\rangle
-\tr  HH^* -|x|^2
\\&=& -\left[ \|H\|^2_\HS +i\sum_{j,k=1}^m
\xi_j|H_{jk}|^2 + \left(1 -\frac i2
\sum_j\xi_j\right)|x|^2\right]
\\&=& -\left[\sum_{j\le
k}\left( 1 +\frac i2 (\xi_j+\xi_k)\right)|\wh H_{jk}|^2  +
 \left(1 - \frac i2 \sum_j\xi_j\right)|x|^2\right]\,,
 \end{eqnarray*}
where 
\begin{equation*}  
\wh H_{jk} = \begin{cases} \sqrt 2 \, H_{jk} &\text{for } j<k
\\  H_{jk} &\text {for } j=k \end{cases}. \end{equation*}
Thus,
\begin{align} \ical_{\ep,\ep'} &=  (-i)^{m(m-1)/2}
c_m'' \label{ical4} \\ & \quad \times \int_{Y'_{2m-q}} \int_{\R^{m}}\De(\la)\,\De(\xi)\,
\prod_{j=1}^m |\la_j|\;e^{i\langle
\la,\xi\rangle}\,\ical(\la,\xi)\, e^ { -\ep \sum\xi_j^2
-\ep'\sum\la_j^2} \,d\xi\, d\la\;, \nonumber
\end{align} where
\begin{eqnarray*} \ical(\la,\xi)&=& \frac 1{\pi^{d_m}} \int_\C
\int_{\sym(m,\C)}\: e^{\Ph
(H,x;\xi)} \,dH\,dx\\ &=& \frac{1}{\prod_{j\le k}\left( 1 +\frac
i2 (\xi_j+\xi_k)\right)} \int_\C \: e^{-\left(1 - \frac i2
\sum_j\xi_j\right)|x|^2}\,dx\\ &=& \frac{\pi}{\left(1 - \frac i2
\sum_j\xi_j\right)\prod_{j\le k}\left( 1 +\frac i2(\xi_j+\xi_k)\right) }.\end{eqnarray*}

To evaluate $\lim_{\ep,\ep'\to 0+}\ical_{\ep,\ep'}$, we first
observe that the map
$$(\ep_1,\dots,\ep_m)\mapsto
\int_{\R^{m}}\De(\xi)\, \;e^{i\langle
\la,\xi\rangle}\,\ical(\la,\xi)\, e^ { -\sum\ep_j \xi_j^2}
\,d\xi$$ is a continuous map from  $[0,+\infty)^m$ to the tempered
distributions.  In addition, since the integrand in \eqref{ical4}
is invariant under identical simultaneous permutations of the
$\xi_j$ and the $\lambda_j$, it follows that the integral  equals
$m!$ times the corresponding integral over $Y_{m-k}$.  Hence, by \eqref{beta-ical} and \eqref{ical4}, we
have
\begin{eqnarray*} b_{0q}(m) &=& \frac
{(-i)^{m(m-1)/2}} {\pi^{2m}\,\prod_{j=1}^{m-1} j!}\;\lim_{\ep'\to
0^+}\;\lim_{\ep_1,\dots,\ep_m\to 0^+}
\int_{Y_{2m-q}}d\la \\ & & \times\  \int_{\R^{m}}
\De(\la)\,\De(\xi)\, \prod_{j=1}^m |\la_j|\;e^{i\langle
\la,\xi\rangle}\,\ical(\la,\xi)\, e^ { - \sum\ep_j\xi_j^2
-\ep'\sum\la_j^2} \,d\xi\;.\end{eqnarray*} Letting
$\ep_1\to 0,\dots,\ep_m\to 0,\ep'\to 0$ sequentially, produces
the desired result.
\end{proof}

\subsection{Evaluating the inner integral} \label{C2}

The last step is to evaluate the inner integral.  
We begin by writing
\begin{equation}\label{kcrit*}b_{0q}(m) = \frac{(-i)^{m(m-1)/2}}{\pi^{2m}\prod_{j=1}^{m-1} j!}
\int_{Y_{2m-q} }\, \prod_{j=1}^m |\la_j|\,\De(\lambda)\,\ical_{\lambda} d\la\;,\end{equation} where
$$\ical_{\la}=\int_{\R^m}\frac {
 \De(\xi)\,e^{i\langle \lambda,\xi\rangle}\,d\xi}{\left(1 - \frac{i}{2}\sum\xi_j\right)\prod_{j\le k}\left[ 1 +\frac i2(\xi_j+\xi_k)\right]}.$$  In order to simplify the formula, we make the
change of variables $\xi_j \to t_j+i$ to obtain
$$
\ical_{\la} =-(-2i)^{\frac {m^2+m+2}2}\, 
e^{-\sum \la_j}\,\ical_{\la,m+2}\,,$$
where
$$ \ical_{\la,c} =
\int_{(\R-i)^m}\frac {
\De(t)\,e^{i\langle \lambda,t\rangle}}
 {\left(\sum t_j+ic\right)\prod_{1\le j \le
k\le m} (t_j+t_k)}\,dt\,.$$
Putting this together we have
\begin{eqnarray}\label{kcrit}b_{0q}(m) = \frac{(-i)^{m^2-1}2^{\frac {m^2+m+2}2}}{\pi^{2m}\prod_{j=1}^{m-1} j!}
\int_{Y_{2m-q} }d\la\, \prod_{j=1}^m |\la_j|\,\De(\lambda)\,
e^{-\sum \la_j}\,\ical_{\la,m+2}\;.\end{eqnarray} 

Now we need the following lemma from \cite{DSZ2} where the authors evaluated the integral using iterated residues to derive the result.

\begin{lem} \label{xiint} Let $0\le p\le m$ and let $c>0$.  Then for
$$\la_1> \cdots >\lambda_p
> 0 > \lambda_{p+1}>\cdots
>\la_m\;,$$
we have
$$
 \int_{(\R-i)^m}\frac {
\De(t)\,e^{i\langle \lambda,t\rangle}}
 {\left(\sum t_j+ic\right)\prod_{1\le j \le
k\le m} (t_j+t_k)}\,dt =\left\{\begin{array}{ll}\displaystyle
i^{m^2-1}\,\frac{\pi^m}{c}\, e^{c\lambda_m} \quad &\mbox{for
}\ p<m\\[10pt]
\displaystyle i^{m^2-1}\,\frac{\pi^m}{c} &\mbox{for }\ p=m
\end{array}\right.\ .$$
\end{lem}
\noindent By setting $p=2m-q$ and $c=m+2$ in the above lemma and substituting this formula into \eqref{kcrit} we obtain the desired result.

\section{Proof of Theorem \ref{Conj}}\label{D1}

From Theorem \ref{b0qtheo} and the definition of $n_q$ we obtain
\begin{align}
n_{q}(m) = \frac{2^{\frac {m^2+m+2}2}}{(m+2)\prod_{j=1}^{m} j!}
\int_{Y_{2m-q} }  d\la & \,
\left| \prod_{j=1}^m\lambda_j \right| \,\De(\lambda)\,e^{
-\sum_{j=1}^m \la_j}  \label{n_qequation} \\ &\times
\begin{cases} e^{(m+2) \lambda_m} & \text{for $q>m$}
\\1 & \text{for $q=m$}
\end{cases} \,. \nonumber
\end{align}
When $q=m$, we can apply Lemma \ref{q=m} directly to the above integral and simplify to obtain
\begin{equation}\label{nmformula}
n_m(m) = 2\,\frac{m+1}{m+2}.
\end{equation}
For $m \leq q \leq 2m$, we can apply Lemma \ref{q>q+1} with $p=2m-q$ and $c=1$ to the integral in \eqref{n_qequation} and therefore
\begin{equation} \label{nqinequality}
n_{q+1}(m) < \left(\frac{2m-q}{2m-q+1}\right)^2 n_{q}(m).
\end{equation} 
By definition $n(m) = \sum_{q=m}^{2m} n_q(m),$ thus it follows from \eqref{nmformula} and \eqref{nqinequality} that
\begin{align*}
n(m) & < \ 2\frac{m+1}{m+2} + 2\frac{m+1}{m+2} \ \sum_{i=m}^{2m-1} \prod_{j=m}^{i} \left(\frac{2m-j}{2m-j+1}\right)^2 \\
& = 2\frac{m+1}{m+2} \left(1 + \sum_{i=m}^{2m-1} \left(\frac{2m-i}{m+1}\right)^2 \right) \\
& = 2\frac{m+1}{m+2} \left(1 + \frac{m(2m+1)}{6(m+1)} \right) \\
& = \frac{2m+3}{3}\,. 
\end{align*}

\subsection*{Acknowledgments} I would like to thank my advisor, S. Zelditch, for his guidance and helpful suggestions.  I would also like to thank H. Hezari for reviewing the manuscript and providing useful comments.

\end{document}